# A new tool for precise mapping of local temperature fields in submicrometer aqueous volumes

Alexey M. Romshin[1], Vadim E. Zeeb[2,*], Artem K. Martyanov[1], Oleg S. Kudryavtsev[1], Dmitrii G. Pasternak[1], Vadim S. Sedov[1], Victor G. Ralchenko[1], Andrey G. Sinogeikin[3], Igor I. Vlasov[1,*]

[1]Prokhorov General Physics Institute of the Russian Academy of Sciences, Vavilov str. 38, Moscow 119991, Russia

[2]Institute of Theoretical and Experimental Biophysics of the Russian Academy of Sciences, Pushchino, Moscow Region 142292, Russia

[3]Wonder Technologies LLC, Skolkovo Innovation Center, Bolshoy blvd.42, Moscow, Russia

[*]- vlasov@nsc.gpi.ru, zeebvad@gmail.com

Nanodiamond hosting temperature-sensing centers constitutes a closed thermodynamic system, with the only window of energy exchange with the environment without direct contacts of a sensor with intracellular substrates, which is in fact the property of an ideal nanosized thermometer. Here, a new design of a nanodiamond thermometer, based on a 500-nm luminescent nanodiamond embedded into the inner channel of a glass submicron pipette is reported. All-optical detection of temperature, based on spectral changes of the emission of "silicon-vacancy" centers with temperature, is used. We demonstrate the applicability of the thermometric tool to the study of temperature distribution near a local heater, placed in an aqueous medium. The calculated and experimental values of temperatures are shown to coincide within the measurement error at gradients up to 20 °C/μm. Until now, temperature measurements on the submicron scale at such high gradients have not been performed. The new thermometric tool opens up unique opportunities to answer the urgent paradigm-shifting questions of cell physiology thermodynamics.

The invention of the Patch Clamp method[1] and development of highly sensitive fluorescent $Ca^{2+}$ indicators for imaging of free calcium concentration in the living cell[2] led to a revolution in our understanding of fundamental mechanisms underlying ionic channels' functioning. These two outstanding methodological achievements established ultralocal experimental control of the cell membrane potential and intracellular free calcium concentration, both parameters being involved in the equation for electrochemical potential [3,4]. Wide usage of this equation in physiology strongly implies that a distinguishing feature of life manifestation is not only highly organized structural changes in space and time in the living cell, but tuned flows of energy as well. The thermodynamic description of energy flows relies on another parameter of the equation for electrochemical potential, temperature; however, its experimental use as an ultralocal variable for precise temperature control in micro- and nano-volumes in a living cell is still limited by imperfections of current methodological approaches as discussed in ref. 5-7, but not at all by principled theoretical limitations of the temperature usage as a macroscopic thermodynamic parameter in nanoscopic watery volumes, where the number of water molecules is still macroscopic.

Here we describe a new approach to practical implementation of ultralocal thermometry based on a specially designed luminescent nanodiamond, which matches well nanoscale thermodynamic requirements in accuracy with no needs for intracellular calibration and allows perfect and easy nanoscale spatial targeting. Only one diamond crystallite is suggested to use as a thermometer, which spractically excludes the toxicity of the thermosensor when applied to living cells. Until now, the diamond thermometry of biological objects was carried out mainly by introducing a large amount of diamond nanoparticles into a cell from a solution [8-11]. The dissimilarity of temperature-sensing characteristics of individual nanoparticles, their random motion and aggregation seriously limit the use of this approach to the precise mapping of local temperature fields in submicrometer volumes. Note that a single diamond crystallite has already been used to measure temperature in a living tissue. However, in those experiments, fairly large diamonds ~100 μm, attached to the end of the fiber, were used[12] and intended to measure some average temperature in the tissue.

It is important to emphasize that the diamond hosting temperature-sensing defects possesses infinite photostability, and principled environmental insensitivity, since it constitutes a closed thermodynamic system, with the only window of energy exchange with environment without direct contacts of temperature sensor with intracellular substrates, which is in fact the property of an ideal nanosized thermometer. This is not the case with fluorescent molecular thermometers [5-7].

The main features of the new design of nanodiamond thermometers are as follows. Temperature measurement with submicron spatial resolution is carried out with one luminescent diamond nanocrystal, which is fixed at the entrance of the inner channel of a micro- (nano-) pipette (Fig. 1). To bind the diamond to the tip of the micropipette, a 5-µL drop of distilled water is applied to the substrate with crystallites grown on it (Fig. 1a). At this time, some of the diamond crystallites pass into the aqueous medium. Then, the submicron pipette touches the surface of the drop, and a column of water is drawn into its inner channel under capillary forces (Fig. 1b). When the diamond particle approaches the pipette, due to the good adhesion of the diamond to the glass walls of the pipette, it is likely to adhere either before entering the capillary (Fig. 1c) or at its inlet (Fig. 1d). To securely bind the diamond crystallite to the pipette, the glass tip is slightly melted with an electric heater. The temperature sensor is precisely positioned in space using a micromanipulator identical to that used in the patch-clamp method. The positioning accuracy is determined by the technical characteristics of the manipulator and in our case is within 50 nm. In this work, we use a 500-nm diamond particle containing luminescent "silicon-vacancy" (SiV) centers. The diamond is produced by a chemical vapor deposition (CVD) technique (see Methods). All-optical detection of temperature is used [13-20], which is based on the dependence of the spectral position of the maximum of the zero-phonon line (ZPL) of SiV luminescence on temperature. The reading and processing of information about the temperature are carried out using a commercial (HORIBA) confocal Raman/luminescence spectrometer and a home-developed software algorithm for determining the temperature by the SiV line position (see Methods).

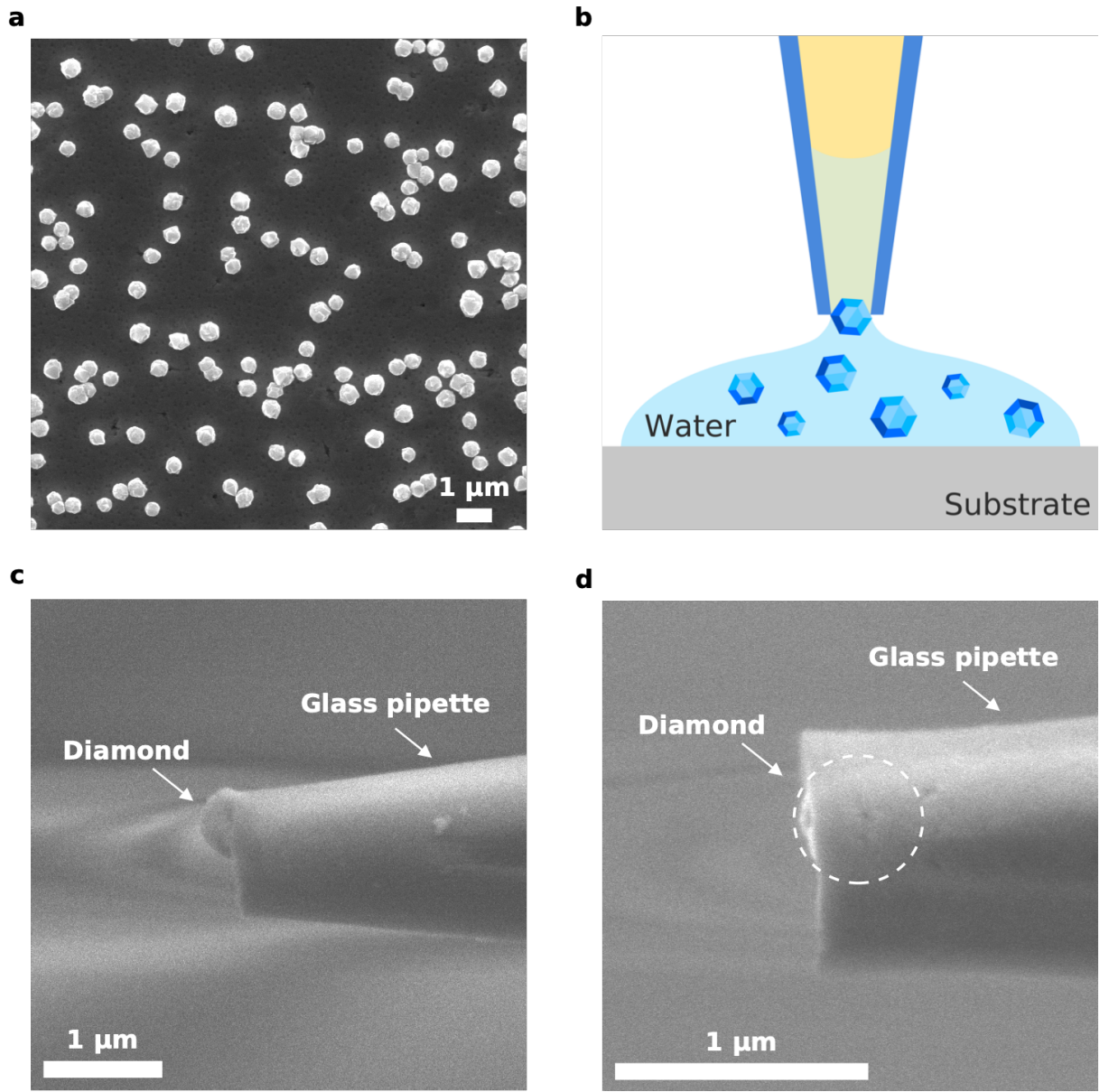

Figure 1. Binding nanodiamond particles to the tip of a glass micropipette: a germanium substrate with diamond crystallites synthesized by CVD technique (a), schematic representation of diamond drawing into the pipette channel from a water drop (b), SEM images of a pipette with a diamond crystallite localized near the entrance (c), or at the inlet (d) into a capillary.

We demonstrate the applicability of the proposed thermometric instrument to the study of temperature distribution near a local heater, placed in an aqueous medium, with a submicron resolution. The heater is constructed in the same way as the thermometer. For this, a small aggregate of 100-nm aluminum nanoparticles is formed at the entrance of the inner channel of another micropipette, which is precisely positioned in space using a micromanipulator (MM). The use of MM to move the thermometer and heater facilitates the selection of a convenient geometry for the experiment. We excite the SiV luminescence in a diamond crystallite and heat the aluminum particles with one laser beam at a wavelength of 473 nm. Both the thermometer and the heater are positioned at a certain angle to the laser beam and mutually perpendicular to each other in a cuvette with water (Fig. 2a). The continuous wave radiation passes through a water-immersion objective (NA = 0.9) and is focused on the aggregate of aluminum nanoparticles, resulting in its heating. The SiV luminescence is excited at the "tail" of the Gaussian intensity distribution of the laser beam. The position of the heater is fixed and does not change throughout the experiment, ensuring a constant temperature gradient near the heater. The temperature dependence on the distance from the heater is measured by moving the thermometer with a step of 0.2-1 μm along the axis of the pipette with the heater. At each step, a photoluminescence (PL) spectrum is recorded (Fig. 2b), the ZPL is approximated by a Lorentzian-shaped curve using the Levenberg-Marquardt method, and its position is determined (see SI). Note that in some works [20] a temperature is determined by SiV line width. Our comparative experiments (not described here) have shown that the "line-width" approach is not as accurate as one used in this work.

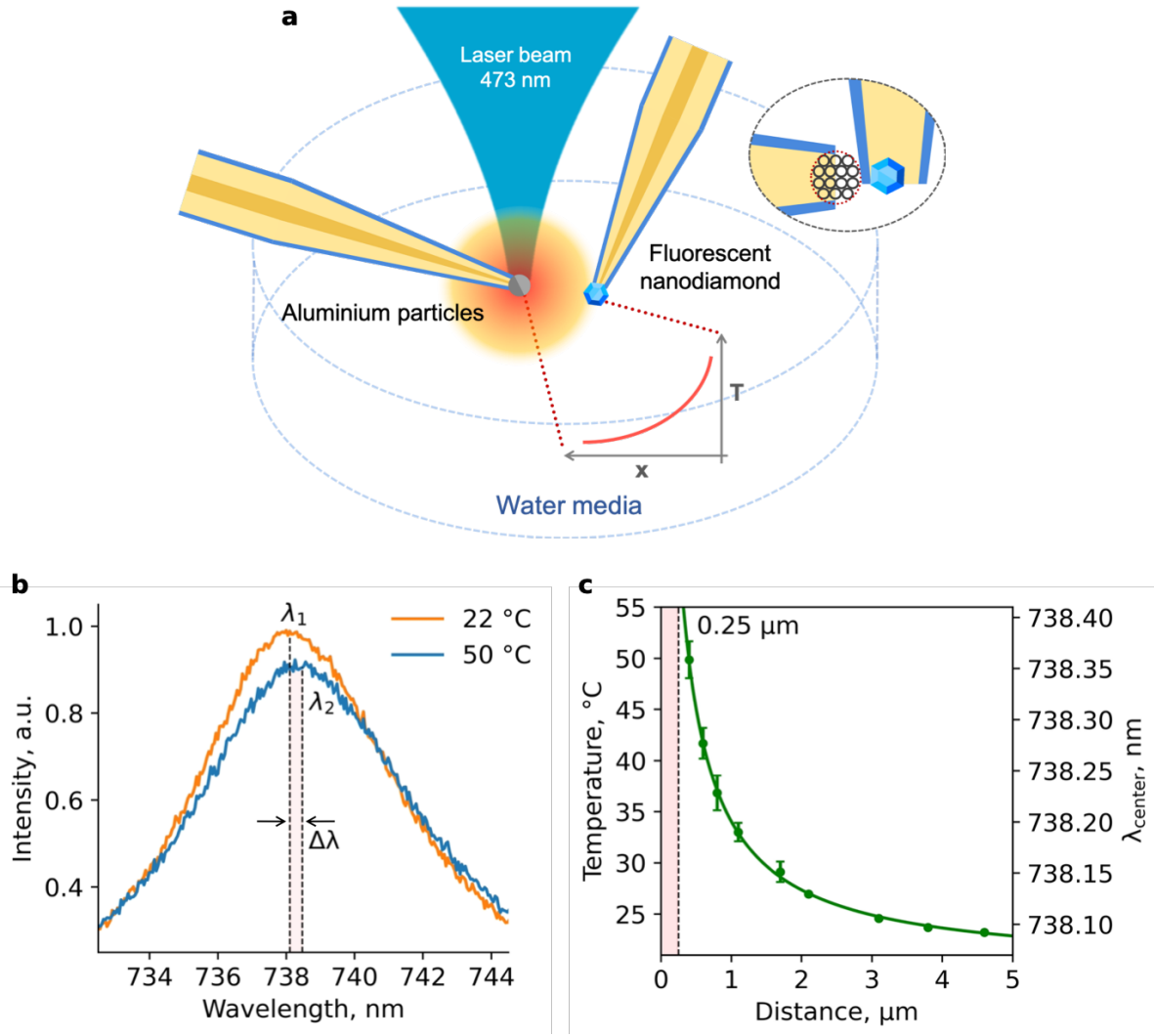

Figure 2. Schematics of temperature distribution evaluation near the local heater: the relative position of the thermometer, heater and laser beam in a cuvette with water (a); PL spectra of the diamond thermometer, measured at different distances from the heater, the positions $\lambda_1$ and $\lambda_2$ of SiV ZPL maxima correspond to 22 °C and 50 °C, respectively (b); dependence of the SiV ZPL position and the temperature on the distance X between the heater surface and the center of the thermometer, dashed line at X=0.25μm corresponds to the distance between the heater surface and the center of the thermometer when they touch each other (c).

Figure 2c shows the dependence of the SiV ZPL position on the distance X between the heater and the thermometer. Pre-calibration of diamond luminescence against the temperature in a thermostat (see SI) allows conversion of the shift in the PL maximum to a local temperature at the particular point. Note the following important features of the experimental data obtained. The 500-nm thermometer:

(1) reproducibly "senses" temperature changes of 2.1 °C degrees over 200 nm (near X=2 μm);

(2) is capable to monitor strong temperature gradients ΔT/ΔX on the submicron scale (the drop ΔT≈15 °C is detected within ΔX≈500 nm near X=1 μm).

When the two pipettes come into contact, the temperature reaches a maximum value of 50.8°C. Hereinafter "X" is the distance between the heater surface and the center of the thermometer and at this point ≈ 0.41 μm, which is a sum of the heater pipette wall thickness of ≈ 0.16 μm and a diamond crystal radius of ≈ 0.25 μm, as shown in the inset to Fig. 2a.

The effect of a diamond nanoparticle, which perturbs the temperature field of the heater, on the accuracy of measuring the spatial temperature distribution is analyzed by numerical methods using the COMSOL Multiphysics software package[21] (comsol.com). The simulated geometry of the "heater-thermometer" system is close to the experimental one. The heater is modeled with an aluminum ball of radius $R_1$ = 400 nm. Its size is determined by the inlet diameter of the micropipette inner channel of 800 nm. The thermometer is modeled with a diamond ball of radius $R_2$ = 250 nm. Both balls are immersed in a water cube with an edge of 40 μm, the center of which coincides with the center of the heater. The boundaries of this cube are thermally insulated. Note that the glass walls of micropipettes can somewhat affect the temperature field of the heater. However, due to the insignificant difference between the thermal conductivities of glass and water, 1.1 and 0.65 W/m·K (ref. 22, 23), respectively, we neglect this effect and ignore the presence of glass micropipettes in our model.

To model the distribution of the temperature field near the heater, it is necessary to know its stationary temperature $T_h$ on the heater surface. Its value depends on the laser power, and in our experiment it is $T_h$≈ 65 °C (see SI). The spatial distribution of the temperature field of a spherical heat source T (X) is derived from the solution of the stationary heat equation in a spherical coordinate system and is described by the following expression:

$$T(X) = T_0 + (T_h-T_0) R_1 / (R_1 + X), \qquad (1)$$

where $T_0$ is the temperature far away from the heater (temperature of the experimental chamber, in our experiment $T_0$ is 22°C), $R_1$ is the radius of the heater, X is the distance to the heater surface. The calculated dependence T (X) is shown in Fig. 3b by the blue curve. Placing the diamond thermometer next to the heater introduces some disturbance to the temperature field around the thermometer, as shown in the temperature distribution map (Fig. 3a).

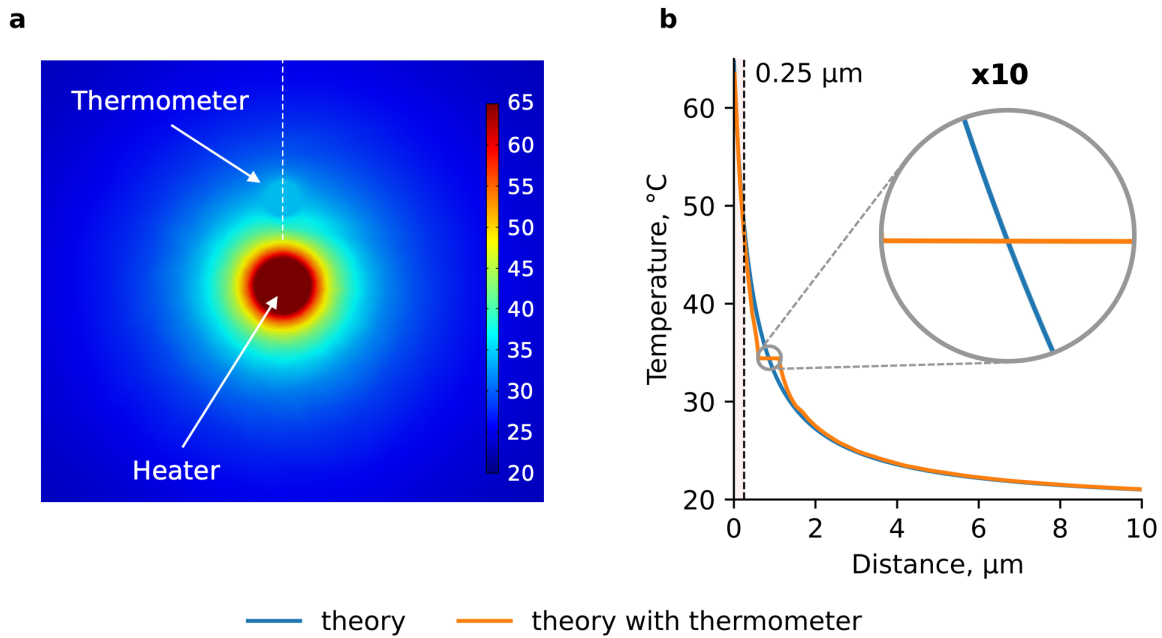

Figure 3. (a) A temperature distribution map in the simulated "heater-thermometer" system; (b) the calculated temperature dependence T (X) without the thermometer (blue curve) and with the thermometer (red curve) the center of which is located at X = 0.85 μm from the heater surface. The T(X) is calculated along a trajectory, shown by a dashed line in (a).

The simulated dependence T (X) in the position of the thermometer at a distance of X = 0.85 μm from the heater is shown in Fig. 3b with a red curve. The temperature is calculated along a trajectory, shown in Fig. 3a. The T (X) distribution deviates from the red curve near the thermometer towards the lower T in front of the thermometer and towards the higher T behind the thermometer. Inside the thermometer the temperature drop is only 0.007°C. The temperature leveling inside the thermometer is explained by the extremely high thermal conductivity of diamond (~20 W/cm·K, three orders of magnitude higher than the thermal conductivity of water). The temperatures at a distance of 0.85 μm, calculated in the presence of diamond and without it, practically coincide. Thus, the remarkable property of diamond, its high thermal conductivity, makes it possible to almost exclude the effect of its size on the accuracy of temperature measurement even at as high temperature gradients as 20 °C/μm.

In Figure 4 the simulated temperature distribution T(x) in the absence (blue curve) and in the presence (orange dots) of a diamond thermometer, and experimental data, taken from Fig. 2c (green dots), are compared. One can see the appearance of a systematic error in the temperature measurements, as the thermometer approaches the heater, and the temperature gradient grows up. This error is associated with the finite size of the thermometer and a strong temperature gradient (>20 °C/μm). The systematic error can be significantly reduced by decreasing the size of the thermometer down to 100 nm (see SI). The difference between calculated and experimental thermometer readings could be explained by the difference between the real shape of the heater and the thermometer from the spherical shape used in our numerical model.

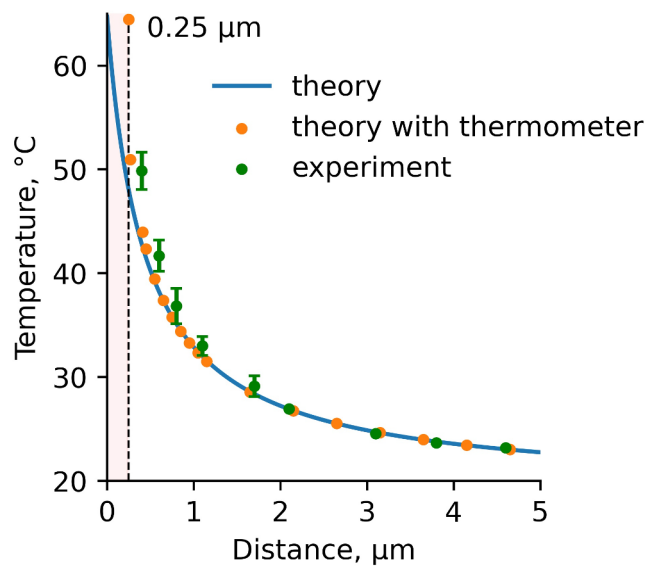

Figure 4. Temperature dependence T (x) without a thermometer (blue curve), calculated thermometer readings (orange dots) and experimental data (green dots).

In conclusion, we develop a new design of a nanodiamond thermometer, which allows precise mapping the temperature fields near local heat sources in submicrometer watery volumes. The use of a high-precision micromanipulator eliminates the limitation of the spatial resolution in temperature inherent for optical methods and determined by the diffraction limit of the optical instrument. In our diamond nanothermometer, unlike its predecessors, only one diamond nanocrystal is used, which has the best luminescent properties for predetermined diamond size: the smallest line width and the highest intensity of SiV luminescence. Such a crystal is selected from a multitude of individual diamond nanoparticles synthesized by the CVD method in the mode of spontaneous diamond nucleation on a substrate. A high accuracy of temperature measurement on the submicron scale is demonstrated for a 500-nm diamond thermometer and a 800-nm aluminum heater immersed in water. The calculated and experimental values of temperatures coincide within the measurement error at gradients up to 20 °C/µm. Until now, temperature measurements on the submicron scale at such high temperature gradients have not been performed.

The new nano-thermometric instrument presented in this work opens up unique opportunities to answer the urgent paradigm-shifting questions of cell physiology thermodynamics, such as:
(1) can open ionic channels be really very hot (tens of degrees °C) in densely packed clusters[24]?
(2) can separate mitochondria be warmer than the surrounding cytoplasm (conundrum of hot mitochondria[25]?
(3) can the local intracellular temperature be really high ("$10^5$ gap issue")[26-27]?
Further, combining a heater and a thermometer in one unit allows one to implement ultralocal hot spot control inside living cells (call it nanoscopic Temperature Clamp Method), which opens unprecedented opportunities for micro- and nano-scale thermal initiation of physiological processes in living cells and modulation of their rates in different intracellular compartments [28,29].

# Acknowledgments

This work was supported in part by the Human Frontier Science Program RGP0047/2018.

# Methods

## I. Growing SiV-luminescent nanodiamonds.

Nanodiamond containing SiV centers is synthesized on a <111> -oriented germanium substrate in a microwave CVD reactor «ARDIS-100» (2.45 GHz). Individual diamond nanoparticles are grown without substrate seeding, using spontaneous nucleation effect for the initiation of the nanoparticle growth [30]. The choice of the substrate material is explained by the weak adhesion between diamond and germanium, which greatly facilitates the transfer of diamond nanoparticles from the substrate surface into an aqueous medium. The synthesis of diamonds is carried out in a hydrogen-methane (96% : 4%) gas mixture, with the addition of 0.1% silane ($SiH_4$), at a substrate temperature of 700-800 °C , a constant pressure of 75 Torr, microwave power of 4.5 kW, and the deposition time of 30 minutes. At this concentration of silane, the maximum SiV luminescence intensity is achieved for individual nanodiamonds. Their characteristic size is 0.5 μm. The content of SiV centers is estimated ~ $10^4$ centers per particle.

## II. Temperature determination.

The temperature of water in the vicinity of the local heater is determined from the position of the SiV luminescence line maximum $\lambda_{max}$. This characteristic of luminescence is sensitive to temperature changes and is proportional to the cube of its value $\Delta\lambda \sim T^3$ [31]. However, for individual diamonds differing in the concentration and spatial distribution of structural defects and SiV centers, the $\lambda_{max}(T)$ may slightly vary from particle to particle. Therefore, for each nanodiamond selected for the thermometer production, a preliminary temperature calibration is carried out using a thermostat in the range of 20 °C - 150 °C in 10 °C steps (SI, Fig. 4S). At each temperature value, a ZPL position is determined by Levenberg-Marquadt approximation method (SI, Fig.3S).

The heater temperature is determined experimentally from the reference boiling point of water. For this, the power of the input laser radiation P is increased until air bubbles appear near the capillary with a heater immersed in water. It is found that the heater temperature $t_{boil}$ = 100 ° C corresponds to $P_0$ = 4.3 mW. Then the power was decreased to $P_1$ = 2.4 mW. Considering that the room water temperature $t_{RW}$ = 22 °C, the heater temperature is estimated as 65 °C.

# Supplementary Information

## I. Designing an aluminum heater.

A powder of aluminum particles with an average size of 50-100 nm and a mass of 2 mg is dispersed in an aqueous medium of 0.5 ml. The formation of a homogeneous aqueous suspension of high-concentration nanoparticles is achieved by cavitation in an ultrasonic bath for one hour. At the next stage, a submicron pipette is brought to the suspension surface and a column of water with aluminum nanoparticles is drawn into the inner channel of the pipette. As a result, an agglomerate of aluminum particles is formed at the entrance to the pipette capillary. SEM image (Fig. 1S) shows that the formed agglomerate is close in shape to a sphere, whose diameter is equal to the capillary size 0.8 μm.

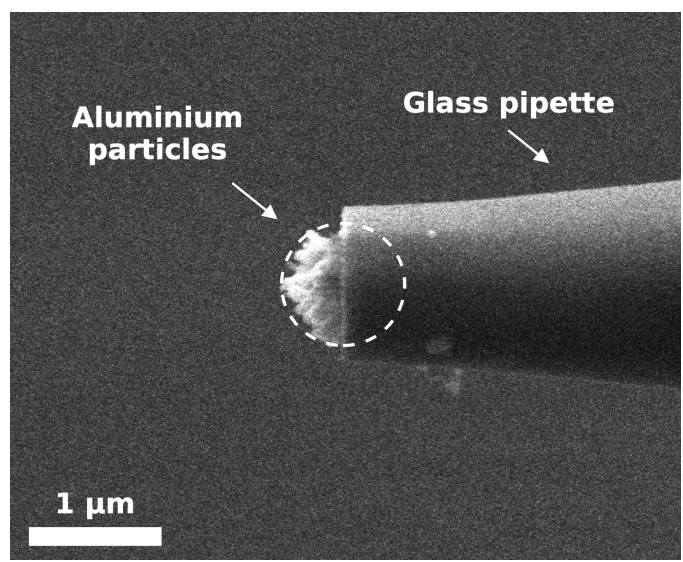

Figure 1S. SEM image of a submicron heater: an agglomerate of aluminum nanoparticles forms at the inlet to the pipette capillary.

## II. Determination of the position of the SiV line.

A typical luminescence spectrum of SiV-containing nanodiamonds is shown in Fig. 2S. At room temperature, the maximum of the ZPL line is located at a wavelength of ≈738 nm. The developed algorithm for determining the line position is based on the approximation of its Lorentzian curve by the Levenberg-Marquadt method. The phonon-related wing, shifted to the long-wavelength region, partially overlaps with the ZPL line, leading to a distortion of its shape. To reduce the effect of such distortion on the approximation precision, we exclude from the approximation the long-wavelength part of the spectrum lying below the level 2/3 of the line intensity maximum $I_0$. To reduce the possible influence of additional sources of luminescence emitting at wavelengths <740 nm, for instance, a small broadband luminescence associated with $sp^2$ carbon on the surface of CVD diamonds, or narrow-band luminescence at 720 nm often accompanying the SiV luminescence[1], we exclude from the approximation the short-wavelength part lying below 1/4 $I_0$. The approximation of the blue line (Fig. 2S) by the Lorentzian curve allows us to estimate the ZPL maximum position with precision $2 \cdot 10^{-3}$ nm, or 0.15 °C on the temperature scale. This parameter can be significantly improved by increasing the SiV line intensity (for example, using separate lasers to excite the SiV and to heat a local volume), the spectral resolution and sensitivity of the luminescence recording system.

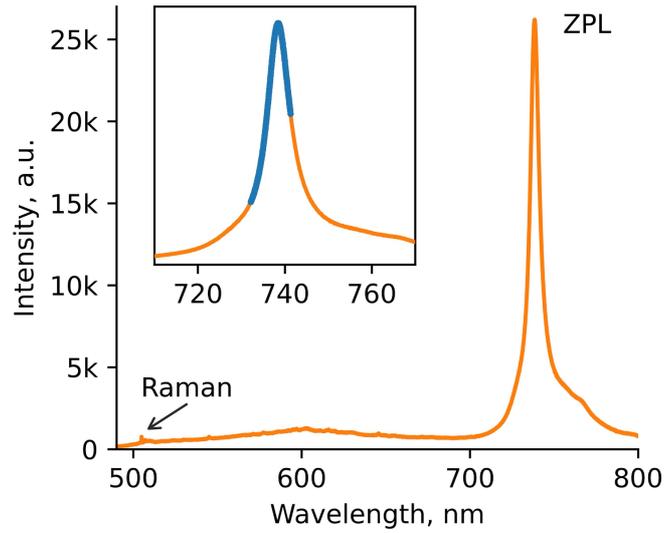

Figure: 2S. Luminescence spectrum of nanodiamonds with SiV centers (orange curve). The part of the spectrum marked in blue in the inset is approximated by the Lorentzian profile.

### III. Temperature calibration of a SiV-luminescent diamond particle.

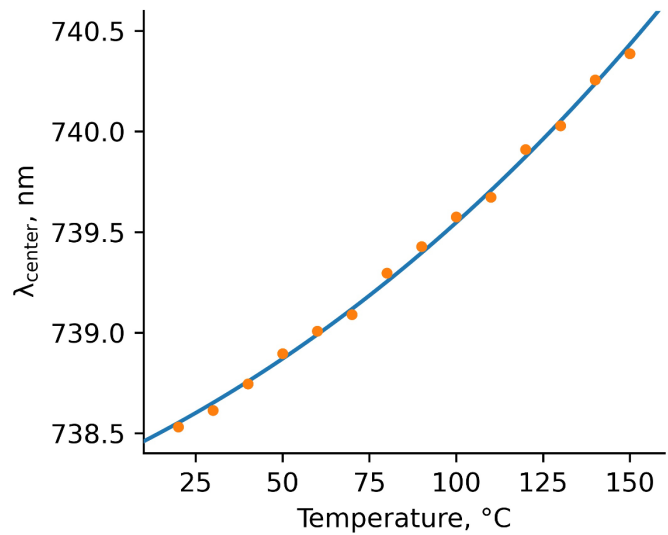

Figure: 3S. Calibration curve of SiV luminescence for a diamond particle attached to the micropipette tip (main text, Figure 1d). $\lambda_{center}(T)$ establishes an unambiguous relationship between the SiV line maximum and temperature. Orange dots - experiment, blue curve - approximation.

**IV. Calculation of T(x) for 100-nm diamond thermometer.**

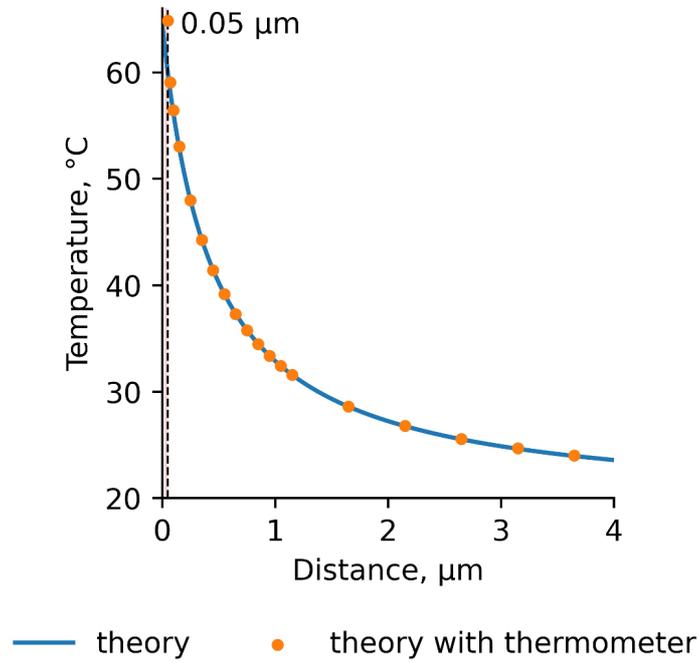

Figure 4S. Calculated temperature dependence T (x) without thermometer (blue curve) and readings of 100-nm diamond thermometer (orange dots).

# References

1. Sedov, V. S., et al. Color Centers in Silicon-Doped Diamond Films. *Journal of Applied Spectroscopy*, **83(2)**, 229-233 (2016).